\documentclass{c2k}

\usepackage{graphicx}

\begin{document}
\begin{titlepage}

\title{Scalable Parallel Implementation of Geant4 Using \\ Commodity
Hardware and Task Oriented Parallel C}

%

\author{Gene Cooperman\inst 1\thanks{Supported in part by NSF Grants 
       CCR-9732330  and ACR-9872114.}, Luis Anchordoqui\inst 2 
       \thanks{Supported by CONICET Argentina.}, Victor Grinberg 
       \inst 1 $^\star$, Thomas McCauley\inst 2 
       \thanks{Supported in part by NSF Grant ACR-9872114.}, 
       Stephen Reucroft\inst 2 $^{\star\star\star}$, and
       John Swain\inst 2 $^{\star\star\star}$
                                }

\institute{College of Computer Science, Northeastern University, Boston, MA 02115,
USA
\and Department of Physics, Northeastern University, Boston, MA 02115,
USA}

\maketitle


\begin{abstract}
 We describe a scalable parallelization of Geant4 using commodity hardware in a
collaborative effort between the College of Computer Science and the Department
of Physics at Northeastern University. The system consists of a Beowulf cluster
of 32 Pentium II processors with 128 MBytes of memory each, connected
via ATM and fast Ethernet. The bulk of the parallelization is done using TOP-C
(Task Oriented Parallel C), software widely used in the computational algebra
community. TOP-C provides a flexible and powerful framework for parallel
algorithm development, is easy to learn, and is available at no cost. Its task
oriented nature allows one to parallelize legacy code while hiding the details
of interprocess communications. Applications include fast interactive
simulation of computationally intensive processes such as electromagnetic
showers. General results motivate wider
applications of TOP-C to other simulation problems as well as to pattern
recognition in high energy physics.
\end{abstract}


\end{titlepage}


\section{Introduction}
\label{sec:introduction}

Among the most CPU-consuming tasks in high energy physics experiments are
the detailed simulations of how detectors respond to high energy particles.
Even today, many physics results are given with contributions to the error
due to the finite amount of Monte Carlo data available, and in many cases
this error is comparable to and even larger than other errors. Even in such
large and well-funded experiments as those at LEP, Monte Carlo statistics
is a large component of the error in precision electroweak measurements.
In addition to its importance in the analysis of data, Monte Carlo simulation
is needed at all stages of the design of experiments in order to understand
and optimize the detector design, as well as to develop a good grasp of the
basic physics issues. In this paper we present the first results of an
ongoing program to use commodity computing to provide parallel computing
for extremely fast Monte Carlo simulations. The aim is to go beyond the
simple event-level parallelism which is commonly used today and actually
run individual events through Geant4 faster than would be possible on any
single workstation or PC. The work has important applications not only
for large scale production, but for the rapid turnaround of ideas and designs
for the working physicist -- the difference between waiting a few minutes
and a few seconds for an event to be simulated and viewed, for example, makes
a world of difference for an interactive user.


\section{Geant4}
\label{sec:Geant4}

Geant4 \cite{GEANTREF,Geant4} is the latest stage in the development of
the GEANT software, superseding the earlier FORTRAN versions with a
new object-oriented approach in C++. For a variety of reasons, in no
small part driven by the wish to work with software which is likely
to see use in the near future, we decided to try to parallelize Geant4.
The aim was to achieve a granularity finer than would be achieved by
simply farming out separate events to separate CPU's and collecting the
results. The approach taken was to perturb the existing software as little
as possible and to modify a section of the code which handles
particle tracking and interaction (a frequent operation) to allow it to
run on multiple CPU's using TOP-C.


\section{Parallelization of Geant4 Using Task-Oriented Parallel C (TOP-C)}
\label{sec:TOPC}

TOP-C (Task Oriented Parallel~C) \cite{TOPC} was initially designed
with the twin goals of easily writing parallel applications and with
the ability to tolerate the high latency typically found on Beowulf
clusters. It is freely available at {\url
{ftp://ftp.ccs.neu.edu/pub/people/gene/topc/}}.  The same application
source code has been run under shared and distributed memory (SMP, IBM
SP-2, NoW, Beowulf cluster).  A sequential TOP-C library is also provided to
ease debugging.  The largest example to date was a
computer construction of Janko's group over three months using
approximately 100 nodes of an IBM SP-2 parallel computer at Cornell
University~\cite{TOPC-SP2}.

    The TOP-C programmer's model~\cite{TOPC} is a master-slave
architecture based on three key concepts:
\begin{enumerate}
\item {\it tasks} in the context of a master/slave architecture;
\item global {\it shared data} with lazy updates; and
\item {\it actions} to be taken after each task.
\end{enumerate}
Task descriptions (task inputs) are generated on the master, and
assigned to a slave.  The slave executes the task and returns the
result to the master. The master may update shared data on all
processes. Such global updates take place on each slave after the
slave completes its current task.  The programmer's model for TOP-C is
graphically described below.

\begin{figure}[htb]\label{topc-fig}
{\footnotesize
\begin{center}\label{diagram}
\setlength{\unitlength}{3.0pt}
\begin{picture}(100,95)
\put(25,90){\makebox(0,0)[b]{\bf MASTER}}
\put(75,90){\makebox(0,0)[b]{\bf SLAVE}}
\put(50,00){\line(0,1){4}}
\multiput(50,15)(0,5){16}{\line(0,1){4}} 
\thicklines \put(5,88){\line(1,0){95}}
\put(20,75){\oval(33,10)}  
\put(20,75){\makebox(0,0){\tt GenerateTaskInput()}}
\put(75,55){\oval(25,10)} 
\put(75,55){\makebox(0,0){\tt DoTask(input)}}
\put(25,32){\oval(45,10)} 
\put(25,32){\makebox(0,0){\parbox[t]{125pt}{\tt
CheckTaskResult(input, \hbox{\ \ \ \ \ \ \ \ \ \ \ \ \ \ \ \ }output)}} }
\put(50,10){\oval(75,10)}
\put(50,10){\makebox(0,0){\tt UpdateSharedData(input, output)}}
\thinlines
\thicklines
\put(34,69){\vector(3,-1){27}} 
\put(36,69){\makebox(0,0)[bl]{\tt input}}
\put(62,51){\vector(-3,-2){18}} 
\put(60,51){\makebox(0,0)[br]{\tt output}} \thinlines
\put(48,37){\vector(3,2){15}} 
\put(48,37){\makebox(0,0)[tl]{(if action == {\tt REDO})}}
\put(25,25){\vector(3,-1){25}} 
\put(36,22){\makebox(0,0)[bl]{(if action == {\tt UPDATE})}}
\end{picture}\break
  \hbox{\large TOP-C Programmer's Model}\break
   \hbox{\large (Life Cycle of a Task)}
\end{center}
}
\end{figure}

The task-oriented approach of TOP-C is ideally suited to parallelizing
legacy applications.  We chose the TOP-C task to be computation of a
particle track in Geant4.  The largest difficulty was in {\it
marshalling} and {\it unmarshalling} the C++ G4track objects that had
to be passed to the slave processes.  Marshalling is the process by
which one produces a representation of an object in a contiguous
buffer suitable for transfer over a network, and unmarshalling is the
inverse process.

We developed a 6-step software methodology to allow ourselves to
incrementally parallelize Geant4, allowing us to isolate
individual issues. The six steps were:
\begin{enumerate}
\item the use of \*.icc (include) files to isolate our code
from the original Geant4 code;
\item collecting the code of the inner loop in a separate
routine, {\tt DoTask()}, whose input was a primary particle track,
and whose output was the primary and its secondary particles;
\item marshalling and unmarshalling the C++ objects for particle tracks
({\it gdb}, a symbolic debugger, and
{\it etags} an emacs facility for a source code browser, were
invaluable here for inspecting the internals of the objects);
\item integrating the marshalled versions of the particle tracks
with the calls to {\tt DoTask()};
\item adding {\tt TOPC\_init()}, {\tt TOPC\_submit\_task\_input}, and
other routines and then testing as the marshalled particle tracks
were sent across the network; 
\item and finally adding {\tt CheckTaskResult()}, which inspected
the task output, and added the secondary tracks to the stack, for
later processing by other slave processes.
\end{enumerate}

Prior to the fifth step, all debugging was in a sequential setting.
The maturity of the TOP-C library then allowed us to create fully
functioning parallel code in less than a day.

\section{A Test Simulation of Extensive Air Showers}
\label{sec:TESTS}

We describe here a simple test of the parallelized Geant4 code described
above. So far we have confined work to electromagnetic calorimetry, one
of the most time-consuming, yet physically understandable tasks to simulate.

With the observation of ultrahigh energy cosmic-ray induced air
showers initiated by primaries carrying over $10^{20}$ eV
\cite{COSMICS,AUGER}, there is a growing interest in better-modelling
particle interactions \cite{hadronic-models}. The currently most
popular programs \cite{AIRES,CORSIKA} use dedicated particle
transport and interaction codes to perform the simulation.
Invariably they contain approximations in order to make the code
run in a reasonable time, but these approximations must at some
point be tested against our best models of physics. In addition,
these programs can lack the flexibility of a general-purpose
program like Geant4. To this end, we consider the modelling of
ultrahigh energy air showers induced by gamma rays, for now taking
into account only electromagnetic interactions. Inclusion of 
hadronic interactions is underway, pending a better understanding
of how to handle them at ultrahigh energies using Geant4.

In the case we consider here, the description of the
calorimeter is moderately complicated. The atmosphere is defined
by a stack of 230 layers
of increasing thickness and decreasing density with the height
above sea level. The layer thicknesses start at 50 m (sea level)
and at higher altitudes are as thick as 1 km. 
The variable density was modeled using Linsley's
parametrization of the U.S. Standard Atmosphere \cite{atmosphere}.

Preliminary comparisons with the serial version of the code show
excellent agreement, and comparisons with other shower simulation codes
are underway.

\section{Conclusions}
\label{sec:CONCLUSIONS}

Geant4 (approximately 100,000 lines of C++ code) was successfully
parallelized using TOP-C.  This was done despite the fact that none of
our group had prior experience with Geant4.  It remains to obtain
timing tests on a long run with many processors.  Initial results
for the example described
indicate that a single task in our example requires approximately 1~ms
of CPU time.  Hence, it will be essential to submit approximately 100
particles for a single slave process to compute, in order to overcome
network overhead.  Optimization of the parallel implementaion is 
underway and we are also interested in collaboration with other
groups who may have needs for the speedups that our methodology
offers.

Task Oriented Parallel~C seems to be well-suited to the problem
of parallelizing Geant4, and would likely be well-suited to other
high energy physics applications as well. Its flexibility and
simplicity makes it possible to envision enormous speedups for Geant4
within a single event, something not often considered in high energy
experiments, but offering many advantages over the usual, trivial
parallelism, especially during interactive data analysis and code or
hardware design.




\begin{thebibliography}{9}
\bibitem{GEANTREF} {\url {http://wwwinfo.cern.ch/asd/geant/ }}
\bibitem{Geant4} {\url {http://wwwinfo.cern.ch/asd/geant4/geant4.html}}
\bibitem{TOPC} G.~Cooperman, ``TOP-C:  A Task-Oriented Parallel~C
Interface'', {\sl $5^{\hbox{th}}$ International Symposium on High
Performance Distributed Computing} (HPDC-5), IEEE Press,  1996,
pp.~141--150.
\bibitem{TOPC-SP2} G.~Cooperman, W.~Lempken, G.~Michler and M.~Weller,
``A New Existence Proof of Janko's Simple Group $J_4$'',  {\sl
Progress In Mathematics}~{\bf 173}, Birkhauser, 1999, pp.~161--175.
\bibitem{COSMICS} S. Yoshida and H. Dai, ``The Extremely High Energy Cosmic Rays'',
J. Phys. G {\bf 24}, 905
(1998).
\bibitem{AUGER} The Pierre Auger Observatory (a surface array plus
an optical air fluorescence detector) is currently under construction.
{\url{http://www.auger.org/ }}
\bibitem{hadronic-models} R. S. Fletcher, T. K. Gaisser, P. Lipari and T.
Stanev, 
Phys. Rev. D {\bf 50}, 5710 (1994); 
N.N. Kalmykov, S.S. Ostapchenko, A.I. Pavlov,
Bull. Russ. Acad. Sci. Phys. {\bf 58}, 1966 (1994); 
J. Ranft, 
{\url{astro-ph/9911232}} at {\url{http://xxx.lanl.gov}}. 
Improvements in the interaction models are also
underway. A preliminary {\sc nexus} skeleton has been already
reported in, H. J. Drescher, M. Hladik, S. Ostapchenko, and K.
Werner, 
{\url{hep-ph/9806407}} -- 
{\url{hep-ph/9806410}} 
at {\url{http://xxx.lanl.gov}}.
\bibitem{AIRES} S. Sciutto, {\it Air Shower Simulations with the AIRES system},
in {\it Proc. XXVI International Cosmic Ray Conference}, (Eds. D.
Kieda, M. Salamon, and B. Dingus, Salt Lake City, Utah, 1999)
vol.1, p.411, {\url{astro-ph/9905185}} at {\url{http://xxx.lanl.gov}}.
\bibitem{CORSIKA} D. Heck {\it et al.}, {\it CORSIKA 
(COsmic Ray Simulation for KASCADE)},
FZKA6019 (Forschungszentrum Karlsruhe) 1998; updated by D. Heck
and J. Knapp, FZKA6097 (Forschungszentrum Karlsruhe) 1998.
\bibitem{atmosphere} {\it U.S. Standard Atmosphere} 1962, updated 1976, U.S. Government
Printing Office.

\end{thebibliography}
\end{document}